\documentclass[11pt]{article}
\usepackage{epsfig,dsfont}

\begin{document}  
\sffamily

\thispagestyle{empty}
\vspace*{15mm}

\begin{center}

{\LARGE 
Coherent center domains in SU(3) gluodynamics \vskip2mm
and their percolation at $T_c$}
\vskip30mm
Christof Gattringer
\vskip8mm
Institut f\"ur Physik, Unversit\"at Graz, \\
Universit\"atsplatz 5, 8010 Graz, Austria 

\end{center}
\vskip30mm

\begin{abstract}
For SU(3) lattice gauge theory we study properties of static quark sources 
represented by local Polyakov loops. We find that for temperatures both below
and above $T_c$  
coherent domains exist where the phases of the local loops have similar 
values in
the vicinity of the center values $0, \pm 2 \pi/3$. The cluster properties of
these domains are studied numerically. We demonstrate that the deconfinement
transition of SU(3) may be characterized by the percolation of suitably
defined clusters.
\end{abstract}
\vskip20mm
\begin{center}
{\it To appear in Physics Letters B}
\end{center}

\newpage
\setcounter{page}{1}
\noindent
{\bf Introductory remarks}
\vskip2mm
\noindent
Confinement and the transition to a deconfining phase at high temperatures are
important, but not yet sufficiently well understood properties of QCD. With the
running and upcoming experiments at the RHIC, LHC and GSI facilities, it is
important to also obtain a deeper theoretical understanding of the mechanisms 
that drive the various transitions in the QCD phase diagram. 

An influential idea is the Svetitsky-Jaffe conjecture \cite{znbreaking}
which states that for pure gluodynamics the critical 
behavior can be described by an effective spin model
in 3 dimensions which is invariant under the center group $\mathds{Z}_3$ (for
gauge group SU(3)). The spin degrees of freedom are related \cite{ploopeff}
to static quark
sources represented by Polyakov loops, which in a lattice regularization are
given by
\begin{equation}
L(\vec{x}) \, = \, 
\mbox{tr}_c \prod_{t=1}^{N} U_4(\vec{x},t) \; . 
\end{equation}
The Polyakov loop $L(\vec{x})$ 
is defined as the ordered product of the SU(3) valued temporal gauge variables
$U_4(\vec{x},t)$ at a fixed spatial position $\vec{x}$, where 
$N$ is the number of lattice points in time direction
and $\mbox{tr}_c$ is the trace over color indices. 
The loop $L(\vec{x})$ thus is a gauge transporter that closes around 
compactified time. Often also the spatially averaged loop $P =  1/V 
\sum_{\vec{x}} L(\vec{x})$ is considered, where $V$ is the spatial volume. 
Due to translational invariance $P$ and $L(\vec{x})$ have the same vacuum
expectation value. 

\begin{figure}[t]
\begin{center}
\includegraphics[width=10.5cm,clip]{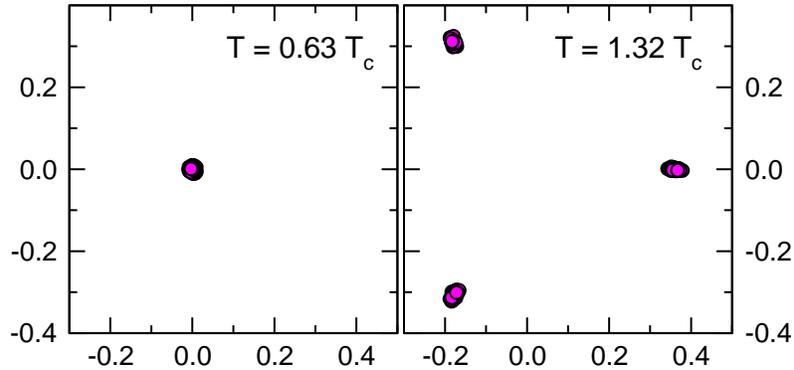}
\end{center}
\caption{Scatter plots of the spatially averaged
Polyakov loop $P$ in the complex plane for
configurations below (lhs.\ panel) and above $T_c$ (rhs.). We show the results
for our $40^3 \times 6$ ensembles. 
\label{ploopscatter}}
\end{figure}

The Polyakov loop corresponds to a static quark source and its vacuum 
expectation
value is (after a suitable renormalization) 
related to the free energy $F_q$ of a single quark, $\langle P \rangle 
\propto \exp(-F_q/T)$, where $T$ is the temperature 
(the Boltzmann constant is set
to 1 in our units). Below the critical temperature $T_c$ quarks are confined and
$F_q$ is infinite, implying $\langle P \rangle = 0$. This is evident in the 
lhs.\ plot of
Fig.~\ref{ploopscatter} where we show scatter
plots of the values of the Polyakov loop $P$ 
in the complex plane for 100 configurations below
(lhs.\ panel) and above $T_c$ (rhs.) \footnote{The numerical results we show 
are from a Monte Carlo simulation of SU(3) lattice gauge theory using the
L\"uscher-Weisz gauge action \cite{LuWe}. We work on various lattice sizes
ranging from $20^3 \times 6$ to $40^3\times 12$. The scale was set \cite{scale}
using the Sommer parameter. In our figures we always use the dimensionless 
ratio $T/T_c$ with the critical temperature $T_c = 296$ MeV calculated for
this action in \cite{TCdet}. All errors we show are statistical errors
determined with a single elimination jackknife analysis.}. In the high
temperature phase quarks become deconfined leading to a finite $F_q$ which gives
rise to a non-vanishing Polyakov loop (rhs.\ in Fig.~\ref{ploopscatter}). 

On a finite lattice, above $T_c$ the phase of the Polyakov loop assumes values 
near the center phases which for SU(3) are $0, \pm 2\pi/3$ (rhs.\ 
plot of Fig.~\ref{ploopscatter}). This is a reflection of the
underlying center symmetry which is a symmetry of the action and the path
integral measure of gluodynamics, that is broken spontaneously above the
deconfinement temperature $T_c$. As long as the volume is finite all three
sectors are populated, while in an infinite volume only one of the three
phase values survives. This center symmetry and its spontaneous breaking are
the basis for the above mentioned Svetitsky-Jaffe conjecture \cite{znbreaking}. 

The relation of the deconfinement transition of SU($N$) gauge theory to 
$\mathds{Z}_N$-symmetric spin models has an interesting implication: For such
spin models it is known that suitably defined clusters made from neighboring
spins that point in the same direction show the onset of percolation at the 
same temperature where the $\mathds{Z}_N$-symmetry is broken spontaneously. 
For, e.g., the Ising system these percolating clusters were identified
\cite{coniglio} as the Fortuin-Kasteleyn clusters \cite{fkclusters}. 
An interesting question is whether the cluster- 
and percolation properties can be directly observed in a lattice simulation of
gluodynamics -- without the intermediate step of the effective spin theory 
\cite{ploopeff} for the Polyakov loops. 

For the case of gauge group SU(2) such cluster structures were
analyzed in a series of papers \cite{satz,fortunato}, while for 
SU(3) the formation of center clusters has not yet been explored. In this paper
we try to close this gap and study the behavior of the local 
loops $L(\vec{x})$ and the possible formation of center clusters. Furthermore,
we study center clusters not only near $T_c$ (where they directly can be 
expected from the Svetitsky-Yaffe conjecture) but explore their emergence and
properties in a window of temperatures ranging from 0.63 $T_c$ to 1.32 $T_c$.   

\begin{figure}[t]
\begin{center}
\includegraphics[width=9cm,clip]{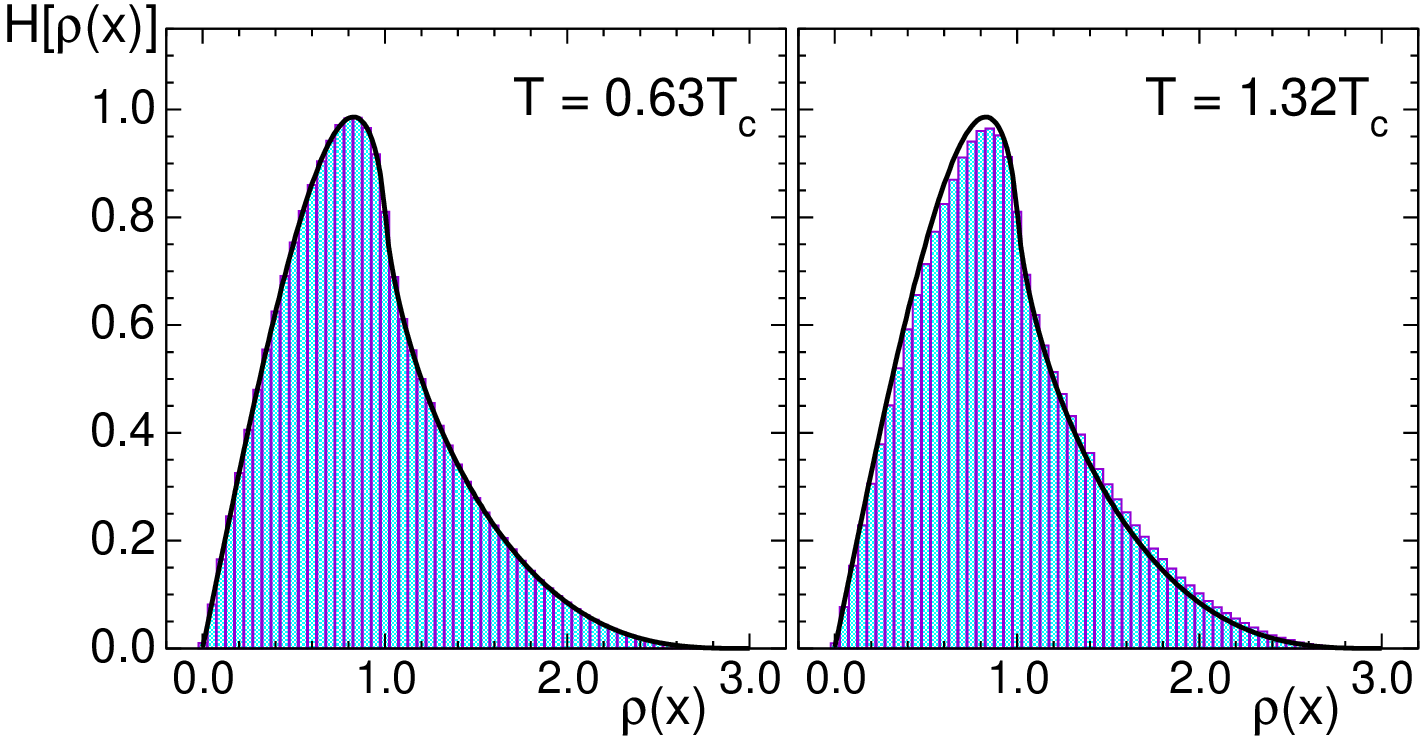}
\end{center}
\caption{Histograms for the distribution of the 
modulus $\rho(\vec{x})$ of
the local loops $L(\vec{x})$ for temperatures below and above $T_c$. The full
curve is the distribution according to Haar measure ($40 \times 6$ ensembles).
\label{rhohistogram}}
\end{figure}

\newpage
\noindent
{\bf Properties of local Polyakov loops}
\vskip2mm
\noindent
For analyzing spatial structures of $L(\vec{x})$ on individual configurations
we write the local loops in terms of a modulus $\rho(\vec{x})$ and 
a phase $\varphi(\vec{x})$,
\begin{equation}
L(\vec{x}) \; = \; \rho(\vec{x}) \, e^{ \, i \, \varphi(\vec{x})} \; .
\end{equation}
The first step of our investigation is to study the behavior of the modulus
$\rho(\vec{x})$. In Fig.~\ref{rhohistogram} we show histograms for the 
distribution of  $\rho(\vec{x})$ in the confined (lhs. plot) and the deconfined 
phase (rhs.). It is obvious, that the distributions of the modulus
$\rho(\vec{x})$ below and above $T_c$ are almost indistinguishable. Furthermore
we find that the distribution follows very closely the distribution 
according to Haar measure, which we show as a full curve. Only above $T_c$ we 
observe a very small 
deviation from the Haar measure distribution. The
Haar measure distribution curves for the modulus and the phase are defined as
\begin{equation}
P(\rho) \; = \; \int D[U] \, \delta(\rho - |\,\mbox{Tr}\,[U]|) \; , \; 
P(\varphi) \; = \; \int D[U] \, \delta(\varphi - \mbox{arg Tr\, }[U]) \; ,
\end{equation} 
where $\delta$ is the Dirac delta-function and $D[U]$ is the Haar integration
measure for group elements $U \in$ SU(3). These two distributions are obtained
from a single group element and thus do not depend on any lattice parameters.

From the fact that the change of the modulus is
very small we conclude that the jump of $\langle P \rangle$ 
at $T_c$, signaling the first order deconfinement
transition, is not driven by a changing modulus of the local loops 
$L(\vec{x})$. 
\begin{figure}[t]
\begin{center}
\hspace*{-8mm}
\includegraphics[width=14cm,clip]{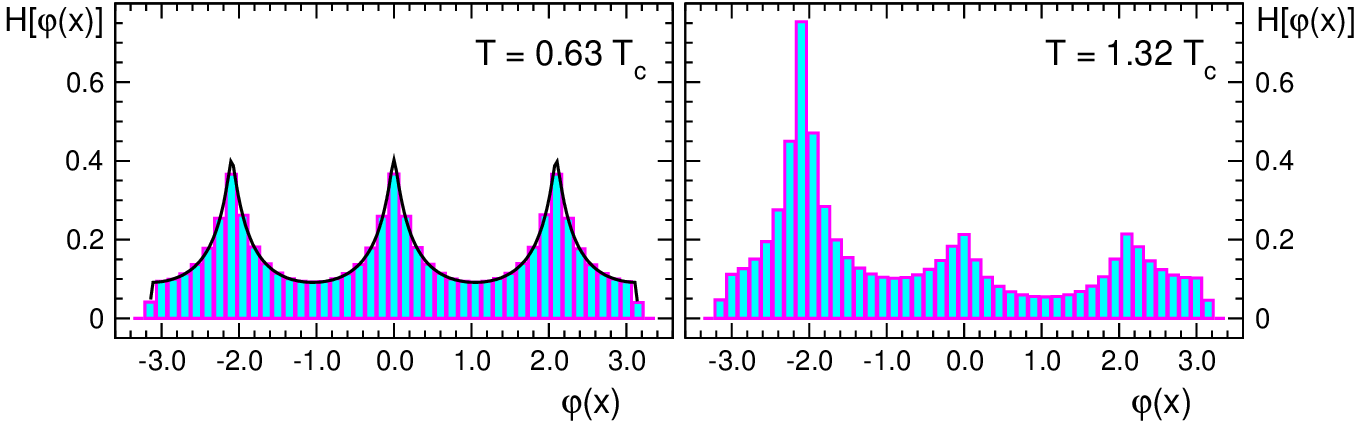}
\end{center}
\caption{Histograms for the distribution of the phase $\varphi(\vec{x})$ of the 
local loops $L(\vec{x})$. We compare the distribution below $T_c$ 
(lhs.\ plot) to 
the distribution in the deconfined phase (rhs.) for the sector of
configurations with phases of 
the averaged loop $P$ near
$-2\pi/3$. The full
curve in the lhs.\ plot is the distribution according to Haar measure
($40 \times 6$ ensembles).
\label{phasehisto}}
\end{figure}
Thus we focus on the behavior of the phase $\varphi(\vec{x})$, and again
study histograms for its distribution. In Fig.~\ref{phasehisto}
we compare the distribution below $T_c$ (lhs.\ plot) 
to the one in the deconfined
phase (rhs.). For the latter we show the distribution for the sector of
configurations
characterized by phases of the averaged Polyakov loop $P$ in the
vicinity of $-2\pi/3$ (compare the rhs.\ of Fig.~\ref{ploopscatter}). 

The distribution of the phases $\varphi(\vec{x})$ is rather interesting:
Also in the confined
low temperature phase (lhs.\ plot in Fig.~\ref{phasehisto}) the distribution 
clearly is peaked at the center phases $-2\pi/3$, $0$ and $+2\pi/3$, and again
perfectly follows the Haar measure distribution (full curve in the lhs.\ plot). 
The distribution is identical around these three phases and the vanishing
result for $\langle P \rangle $ below $T_c$ comes from a phase average, 
$1 + e^{i2\pi/3} + e^{-i2\pi/3} = 0$. 

Above $T_c$ (rhs.\ plot in Fig.~\ref{phasehisto}) the distribution singles out
one of the phases. In our case, where configurations in the sector with phases 
of the averaged loop $P$ near $-2\pi/3$ are used for the plot, it is 
the value $-2\pi/3$ which is singled out. For configurations in one of the other
two sectors (see rhs.\ plot of Fig.~\ref{ploopscatter}) 
the distribution is shifted periodically by $\pm 2\pi/3$. Obviously,
above $T_c$ the distribution is not equal for the three center phases and the
cancellation of phases does no longer work, resulting in a non-zero $\langle
P \rangle$. 

\begin{figure}[t]
\begin{center}
\includegraphics[width=8.9cm,clip]{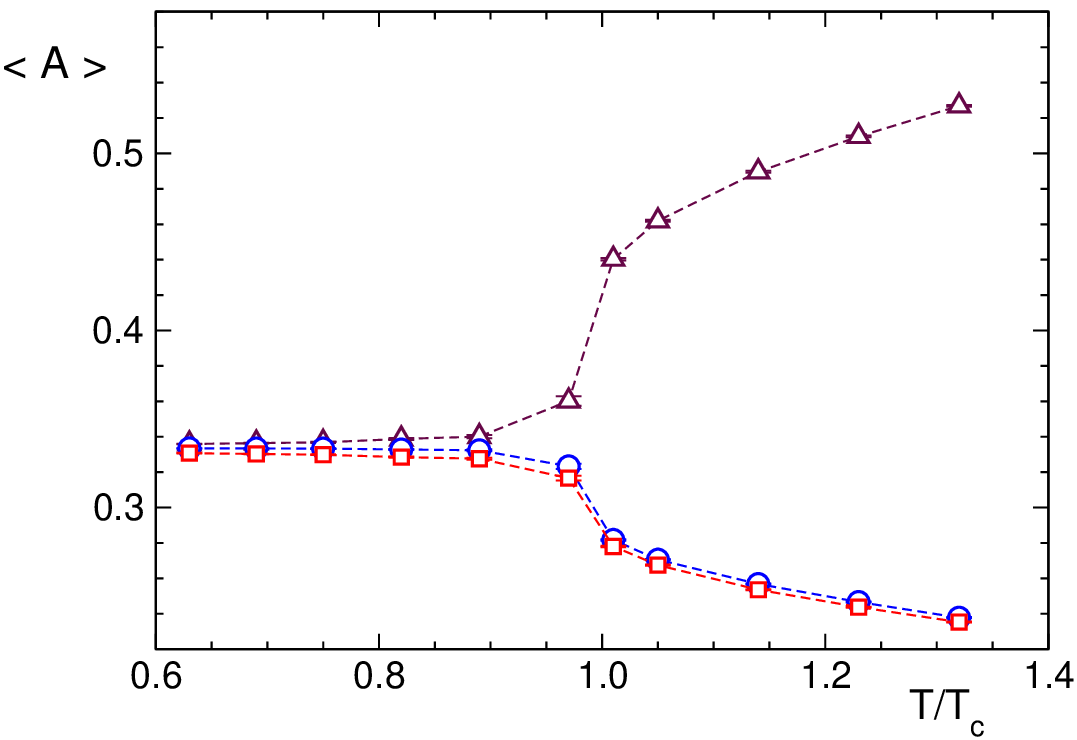}
\end{center}
\caption{Abundance of lattice points $\vec{x}$ with phases of the local loop
$L(\vec{x})$ in the dominant (triangles) and subdominant  
center sectors (circles, squares) as a function of temperature
($40 \times 6$ ensembles). 
\label{abundance}}
\end{figure}

The histograms for the phases $\varphi(\vec{x})$ suggest that at the critical
temperature the local loops $L(\vec{x})$ start to favor phases near one
spontaneously selected 
center value, while phases near the other two center values are depleted.
This is illustrated in more detail in Fig.~\ref{abundance} where we show the
abundance $A$ of lattice points with phases of $L(\vec{x})$ 
near the dominant and 
subdominant center values. 
To define the abundance $A$ we cut the interval $(-\pi, \pi)$ at the 
minima of the distribution of Fig.~\ref{phasehisto} into the
three sub-intervals $(-\pi,-\pi/3)$, $(-\pi/3,\pi/3)$, $(\pi/3, \pi)$, which we
refer to as ``center sectors''. We count the number of lattice points with
phases in each of the three center sectors and obtain their abundance $A$ by
normalizing these counts with the volume.
Fig.~\ref{abundance} shows that at low temperatures
all three center sectors are populated with probability 1/3. Near
$T_c$ one of the sectors starts to dominate while the other sectors are
depleted.

\vskip5mm
\noindent
{\bf Coherent center domains}
\vskip2mm
\noindent
We have demonstrated for a wide range of temperatures
that the center sectors play an important role for
the phases $\varphi(\vec{x})$ of the local loops $L(\vec{x})$, which cluster 
near the center phases
$0, \pm 2\pi/3$ at all temperatures. The
deconfinement transition is manifest in the onset of a dominance of one 
spontaneously selected
center sector. We now analyze whether the values of the 
phases $\varphi(\vec{x})$ are distributed homogeneously in space, or if instead
there exist spatial domains with coherent phase values in the same sector.

In order to study such domains, we use sub-intervals that divide the interval 
$(-\pi,\pi)$ for
the values of the $\varphi(\vec{x})$. For a more general
analysis we introduce the cutting parameter $\delta \geq 0$ and 
define the three 
sub-intervals as $(-\pi + \delta, -\pi/3 -\delta)$, 
$(-\pi/3 + \delta, \pi/3 -\delta)$ and $(\pi/3 + \delta, \pi -\delta)$ 
(which we again refer to as ``center sectors''). For
$\delta = 0$ we obtain the old sub-intervals, while a value of $\delta > 0$ 
allows
us to cut out lattice points where the phases are near the minima of the
distributions shown in Fig.~\ref{phasehisto}. 

\begin{figure}[t]
\begin{center}
\includegraphics[width=9cm,clip]{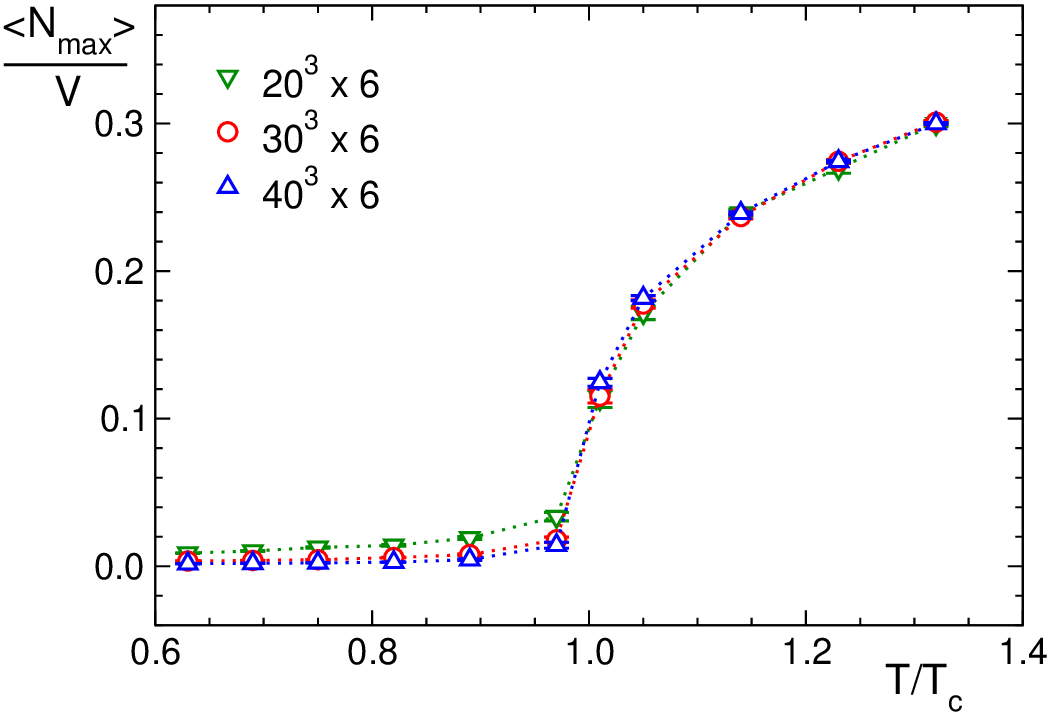}
\end{center}
\caption{Weight of the largest center cluster as a function of
temperature. 
\label{maxcluster}}
\end{figure}

\begin{figure}[t]
\begin{center}
\includegraphics[width=9cm,clip]{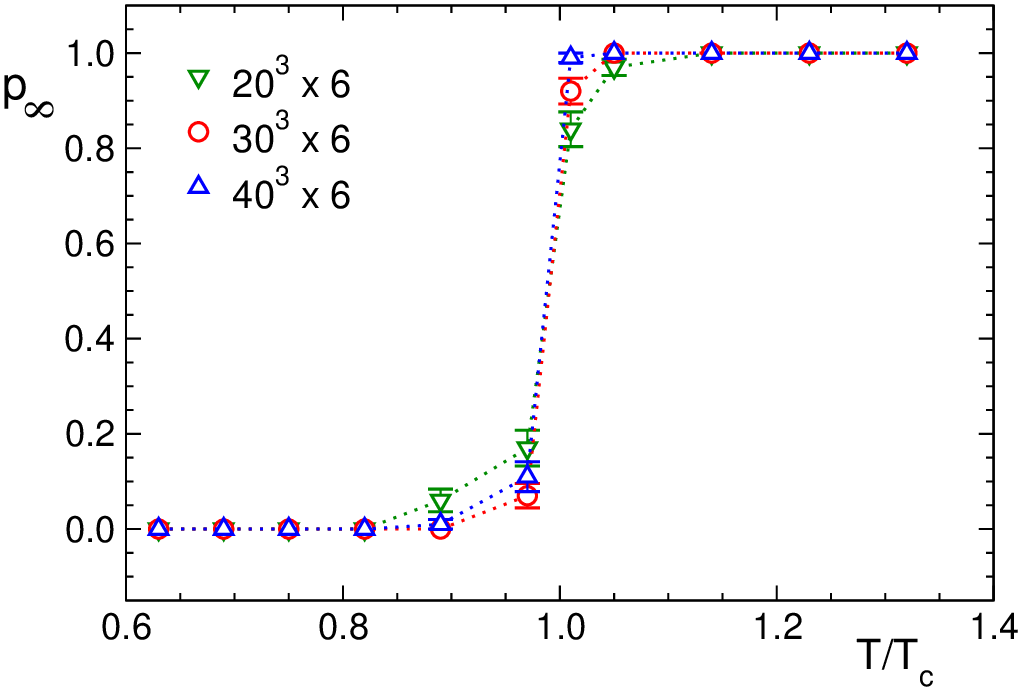}
\end{center}
\caption{Percolation probability $p_\infty$ of the dominant 
center clusters as a function of temperature.}
\label{percplot}
\end{figure}

The definition of the clusters slightly differs from those that have been 
used for the analysis \cite{satz,fortunato} 
of clusters in SU(2) gauge theory. Besides modifications of the 
Fortuin-Kasteleyn prescription studied in \cite{satz}, 
in \cite{fortunato} the bonding
probability between neighboring sites with same sign Polyakov 
loops\footnote{For SU(2) the $L(\vec{x})$ are real and the center phases are
either $+1$ or $-1$.} was introduced as a free parameter. This parameter
could then be tuned such that the onset of percolation agrees with the
deconfinement temperature. In our definition the parameter 
$\delta$ allows one to reduce the lattice to a skeleton of points with 
phases close to the center elements (in intervals of width 
$2(\pi/3 - \delta)$ around the center values). For the plots shown in 
Figs.~\ref{maxcluster} and \ref{percplot} we choose $\delta$
such that roughly those 40 \% of lattice points are cut
where the phases do not strongly lean towards one of the center values. 
We found that near $T_c$ the critical properties of the clusters
(behavior of largest cluster and percolation) are stable when
$\delta$ is varied in a small interval around that value \cite{inprep}
(compare also the discussion in \cite{fortunato}). For example the curve for 
the weight of the largest cluster (see Fig.\ \ref{maxcluster} below, 
where we show a comparison of different spatial volumes for a cut of
39 \%) is form-invariant in a range of cuts from 30 \% to 45 \% and only is
rescaled by a change of the amplitude of less than 15 \%.

In a next step we define clusters by  assigning neighboring lattice sites with
phases $\varphi(\vec{x})$ in the same center sector to the same cluster.   
Once these center clusters are defined we can study their properties and
behavior with temperature using concepts
developed for the percolation problem \cite{percolation}. In
Fig.~\ref{maxcluster} we show the weight (i.e., the number of sites) 
of the largest cluster as a function of
the temperature. For low
temperatures all clusters are small, while as $T$ is increased towards the
deconfinement temperature the largest cluster starts to grow
quickly and above $T_c$ scales with the volume. This property indicates
that in the deconfined phase the system has developed a percolating cluster.
The onset of percolation at $T_c$ is confirmed in Fig.~\ref{percplot} where we
show the percolation probability $p_\infty$ as a function of $T/T_c$. The
percolation probability is computed by averaging an observable which is 1 if a
spanning cluster exists and 0 otherwise. In our case, where we have periodic
spatial boundary conditions, a spanning cluster is defined as a cluster 
who has at least one member site in every $y$-$z$ plane. In other words,
we analyze
percolation in $x$ direction, which is, however, no loss of generality as we
have invariance under discrete spatial rotations. Varying $\delta$ in a range
where the number of points we cut varies between 30 and 45 \% 
(Fig.~\ref{percplot} is for 39 \%), slightly roundens
the transition curve, but leaves the onset of percolation unchanged 
at $T/T_c = 1$.  

An interesting question is the size of the clusters in physical units in the
confining phase, which could be related to some hadronic scale (see also the
discussion in the next section). In order to study this cluster size below
$T_c$ we computed 2-point correlation functions of points within the individual
clusters. These correlators decay exponentially $\propto \exp(- r/\xi)$ 
with distance $r$, and the factor $\xi$ defines a linear size 
of the clusters in lattice units. 
We then analyze $d \equiv 2 \, \xi \, a$, which gives a
definition of the cluster diameter in physical units ($a$ is the lattice
spacing in fm). We find that up to $T = 0.85 T_c$ this diameter is essentially
independent of the temperature, with a value of $d = 0.46(5)$ fm at a cut
of 39 \% and  $d = 0.62(7)$ fm at a cut of 30 \%. Compared to the
expected sizes of rougly 0.5 fm for heavy quark mesons this is a quite
reasonable result for the linear scale of the clusters 
which suggests that the physical role of the clusters below $T_c$
should be studied in more detail (see \cite{inprep}).

\vskip5mm
\noindent
{\bf Summary and discussion}
\vskip2mm
\noindent
We have explored the clustering of the phases 
$\varphi(\vec{x})$ of the local quark sources $L(\vec{x})$ near the center
values, both below and above $T_c$. We find that in the range of temperatures
we consider, $T = 0.63\, T_c$ to $T = 1.32\, T_c$, 
the local Polyakov loop phases
prefer values near the center values and corresponding clusters may be
identified for these temperatures. Using the parameter $\delta$ we can
construct clusters such that the deconfinement transition is characterized by
percolation of the clusters in the dominant sector. 

From the cluster properties a simple qualitative 
picture for confinement and the
deconfinement transition emerges. Below $T_c$ the clusters of lattice points
which have the same
center phase information are small. Only if a quark- and an anti-quark source 
are sufficiently close
to each other they fit into the same cluster and 
can have a non-vanishing expectation value. 
Sources at distances larger than a typical cluster size receive the
independent center fluctuations from different clusters and the correlator
averages to zero. Above $T_c$ the clusters percolate and coherent center
information is available also for larger distances allowing for non-vanishing
correlation at large separation of the sources. 
In this picture deconfinement is a direct 
consequence of a percolating center cluster.

A possible role of local center structures for 
confinement has been addressed also in a different approach,
using a projection of the link variables
$U_\mu(\vec{x},t)$ at all points in space and time to a center element after
fixing to a suitable gauge (see, e.g., \cite{centervortices} 
for a selection of recent results). 
This analysis is motivated by understanding the role of topological
objects for the QCD phase transition. It would be highly interesting to 
study a possible connection of the percolation aspects of the transition 
to the dynamics of such topological objects. Of particular relavance would be
an analysis of a possible relation to calorons which induce strong local
variations of the Polyakov loop that might play an important role 
in the formation of the center clusters \cite{calorons}.

We conclude with a few comments on the extension of the center domain picture 
to the case of full QCD: The fermion determinant describing the dynamical
quarks can be expressed as a sum over closed loops, which may be viewed as
generalized Polyakov loops and are sensitive to the center properties
of the gauge fields \cite{candet}. 
The fermion determinant breaks the center symmetry explicitly and 
acts like an external magnetic field which favors the real sector (phase 0) 
for the Polyakov loop $P$. However, preliminary numerical results 
with dynamical fermions \cite{inprep} show that locally also the 
two complex sectors 
(phases $\pm 2\pi/3$) remain populated. The
corresponding clusters will again lead to a coherent phase information 
for sufficiently close quark lines. As for the pure gauge theory studied in
this letter, the
preliminary results \cite{inprep} show that the 
transition to confinement is again
accompanied by a pronounced increase of the abundance 
for the dominant (i.e.,
real) sector. However, the explicit symmetry breaking 
through the determinant leads to a
crossover type of behavior in the dynamical case. An interesting related 
question, which has already been raised in the literature \cite{chiral}, 
is whether also the chiral transition may be characterized as a percolation 
phenomenon. 
\\
\\
{\bf Acknowledgments:} The author thanks Mike Creutz, Julia Danzer, 
Mitja Diakonov, Christian Lang, 
Ludovit Liptak, Axel Maas, Stefan Olejnik,
Alexander Schmidt and Andreas Wipf 
for valuable comments. The numerical calculations were 
done at the ZID clusters of the University Graz.

\clearpage

\end{document}